# Privacy Preserving Machine Learning for Behavioral Authentication Systems

Md Morshedul Islam
Concordia University of Edmonton
Edmonton, Alberta, Canada
mdmorshedul.islam@concordia.ab.ca

Md Abdur Rafiq
Concordia University of Edmonton
Edmonton, Alberta
mrafiq@student.concordia.ab.ca

## ABSTRACT

A behavioral authentication (BA) system uses the behavioral characteristics of users to verify their identity claims. A BA verification algorithm can be constructed by training a neural network (NN) classifier on users' profiles. The trained NN model classifies the presented verification data, and if the classification matches the claimed identity, the verification algorithm accepts the claim. This classification-based approach removes the need to maintain a profile database. However, similar to other NN architectures, the NN classifier of the BA system is vulnerable to privacy attacks. To protect the privacy of training and test data used in an NN different techniques are widely used. In this paper, our focus is on a non-crypto-based approach, and we used random projection (RP) to ensure data privacy in an NN model. RP is a distance-preserving transformation based on a random matrix. Before sharing the profiles with the verifier, users will transform their profiles by RP and keep their matrices secret. To reduce the computation load in RP, we use sparse random projection, which is very effective for low-compute devices. Along with correctness and security properties, our system can ensure the changeability property of the BA system. We also introduce an ML-based privacy attack, and our proposed system is robust against this and other privacy and security attacks. We implemented our approach on three existing behavioral BA systems and achieved a below 2.0% FRR and a below 1.0% FAR rate. Moreover, the machine learning-based privacy attacker can only recover below 3.0% to 12.0% of features from a portion of the projected profiles. However, these recovered features are not sufficient to know details about the users' behavioral pattern or to be used in a subsequent attack. Our approach is general and can be used in other NN-based BA systems as well as in traditional biometric systems.

## 1 INTRODUCTION

Over the last few years, the world has witnessed the beginning of a revolution taking shape in the field of technology.The smartphone is one of the most significant technological advancements. Recent studies have demonstrated the feasibility of extracting behavioral data, such as touch dynamics, keystroke dynamics, and gait recognition, utilizing smartphone sensors and peripherals. A *Behavioral Authentication (BA)* system [15, 19, 29, 33] utilizes this captured behavioral data for users' authentication. When users use their smart devices, the BA system collects behavioral data from the devices, creates profiles, and sends them to a verifier. The verifier will either store the profiles or use them for verification decisions. Typically, a verifier is an online server that renders real-time verification decisions for a verification claim. In a multi-factor authentication system, BA systems have been used to strengthen other factors. Additionally, BA systems offer a number of alluring features, including continuous authentication [23] and security against "credential sharing" without the requirement for additional hardwares.

A profile in a BA system consists of a set of $m$ $d$-dimensional vectors over real numbers $\mathbb{R}$, each dimension corresponding to a *feature*, and each vector corresponds to a measurement of behavioral features. A database inside the verifier store the profiles created during a trusted registration phase. A tuple consisting of (i) an identity and (ii) a measurement of $n$, ($n < m$) behavioral measurement samples makes up a verification request. The BA system will use a *verification algorithm* to measure the distance between the profile of the claimed identity and the presented data and accept or reject the verification claim. A distance measure is employed in a conventional verification technique to assess the "closeness" of the supplied verification data with the set of samples in the claimed profile. This requires the server to keep a database of user profiles, which will need protection and secure data management because of the security and privacy sensitivity of the profile data.

To train a classifier using profile data, such as a machine learning classifier, where each user is assigned to one class, is an appealing method of verification without the requirement for the server to maintain the profile database. The trained classifier will then be used to evaluate verification requests by classifying the verification data. Recently, neural network (NN) based techniques have been developed to classify face data [28], fingerprint data [27], mouse movements [9], gait data [22], and keystroke data [12]. In each case, the NN classifier creates a parametric function using nodes of each network's layers to learn the mapping between the vector in *training data set* and a set of classes, each corresponding to a user. The output of a NN classifier with $N$ classes on each data sample (vector) of the BA system is an $N$ dimensional probability vector (sum of components equal to 1.0) that is called the *prediction vector*. The classifier's "belief" that the input belongs to the $i$-th class is represented by the $i$-th component of the prediction vector. The final decision of the BA verification algorithm will be based on the prediction vectors of $n$ data samples. This approach removes the need to store the profile data and significantly improves the system's performance. The benefit of using an NN classifier for BA verification is that partial verification judgments can be made for each sample. This can be used for continuous authentication, which regularly verifies the user's identity throughout the authentication session.

**Privacy.** Profile data carries sensitive personal information that must be protected from privacy attackers. A privacy attacker in a BA system can be the verifier or any other external actor. Ideally, a privacy attacker should not be able to access the profile data during the registration and verification phase. In addition, for a given query, the NN classifier's output should not contain any information regarding the training profiles. In reality, however, the leaked information regarding the behavioral data is used to learn about users' behavior, interests for marketing and advertising, or track them across the websites. Biometric profiles are extremely sensitive from a privacy standpoint, and user identities are uniquely recognizable by their profiles. Behavioral profiles can also reveal users' private information, such as their health conditions, physical abilities, or users' skills and behavior patterns, as well as the pattern of usage of applications and devices. Richer profiles (i.e., more behavior data) lead to higher accuracy in authentication, and this provides an incentive to employ more user data and "to keep it around for a longer period of time" [6] which leads to more privacy risk.

Privacy-Preserving Machine Learning (PPML) [3] is a step-by-step approach to preventing data leaks in machine learning. PPML research has exploded recently, and several methods are employed to protect data privacy. Some frequently used techniques in PPML include differential privacy [13], homomorphic encryption [14], and multi-party computation [25]. To protect the profiles' privacy against the privacy attacker of an BA system, an immediate solution is to train the NN classifier for the BA verification algorithm with the encrypted profiles. Then design the verification algorithm as a computation in the encrypted domain using homomorphic encryption to address the verification claim. However, as the BA system typically leverages smartphone devices at the user end to collect data, our interest is to use a non-crypto-based approach to reduce the computation cost at the user end. Moreover, the *changeability* property, which gives the user the ability to the users to refresh a user profile while maintaining correctness and security, is not easily achievable by homomorphic encryption. Additionally, the classifier's output should not provide any information about the model or model's data.

**Our Work.** In [32], *random projection* of profile data was proposed to provide profile privacy for biometric data. The authors demonstrated this by employing a random transformation matrix whose elements are generated using a Gaussian distribution and performing dimension reduction such that the correctness and security of the verification algorithm are maintained. The authors also showed that the approach allows *changeability* of the profile, which is a desirable security property. In [30], the authors extended the work for the behavioral profile and used the discrete distribution of the random matrix to reduce the computation load. Our work strengthens and extends their work in a number of ways. The inclusion of the NN classifier with the verification algorithm makes the privacy problem more challenging, and we solved it with the non-crypto-based approach. Moreover, to limit the information leak, we aggregate the output results suggested in [2] and then publish them as a single output. We also introduced an ML-based privacy attacker for our system.

**Challenges.** The challenges that we face in proposing a privacy-preserving BA system are: (i) constructing and training the BA classifier on projected data by ensuring system correctness, security, and changeability properties; and (ii) protecting the system from different security and privacy attacks. To ensure the correctness and security properties of the system, the transformation needs to preserve the pair-wise distance. Not all transformations do preserve the pair-wise distance. Additionally, users should be able to quickly change their credentials if necessary. From a privacy point of view, the transformation needs to be irreversible, and projected profiles should not carry any information about the features of the plain profiles. Additionally, the transformation should have a low computational overhead so that users can apply it to their smartphone device to transform the profiles.

**Our Approach.** Here are our high-level steps for the BA systems that protect privacy:

– Transform a plain profile into a projected profile by random projection (RP). A random matrix that serves as a secret key will create this projection at the users' end. To project the profile, each user will have a secret random matrix. The projected profile will then be sent from the user end to the verifier. A user will use the same random matrix in RP for both the registration and verification phase of the BA system.
– The server will have $N$ predicted profiles from $N$ users after the registration process. These projected profiles will be used by the server to train an ML classifier. For a query vector of the verification data, the classifier will output the prediction vector. For a verification claim with $m$ query vectors, there are $m$ prediction vectors, and the server will aggregate the prediction vectors to output an accept or reject decision.
– To validate the privacy of the BA system, we will build an NN-based attack model to recover the plain profiles from the projected profiles. To train the attack model, we will use the attack profiles. The profile generator of a BA system is usually publicly available software. By using this profile generator, we will gather behavioral data from users whose data were not used to train the BA classifier.

**Experimental results.** We implemented and evaluated our proposed approach on three different behavioral data sets. The first two have swipe and voice pattern data [18]. The third one has the drawing pattern data [19]. We divided all BA profiles in each data set into two groups: the first group of profiles is used to train the privacy-preserving BA classifier, and the second group of profiles is used as attack data (profiles).

We first used RP to project each profile of Group 1 in each data set. All profiles of swipe data are projected from dimensions 33 to 30, all voice profiles from 104 to 94, and all drawing pattern profiles from 65 to 56. We then used those projected profiles to train the BA classifiers. Each BA classifier has 3-4 stacks of dense layers along with their activation functions and a softmax function layer in the last layer to output a prediction vector. During the training session, within a reasonable number of epochs, all three BA classifiers achieved above 98.0% classification accuracy on the projected data, which is equivalent to 2.0% FRR. However, if the profile is projected by the incorrect random matrix during the verification parse, the classification accuracy is reduced to below 1.0% which is equivalent to 1.0% FRR. The NN-based attack model also has 4-5 stacks of dense layers along with their activation functions. In order to keep the output in the range [0,1], we add a sigmoid function layer in the last layer of each NN. To train the attack model, we projected attack profiles using the random matrices by considering two cases: (i) in case 1, the attacker only knows the distribution of the random matrix, and (ii) in case 2, the attacker knows the secret random matrix of the users. This second scenario corresponds to the most extreme scenario, in which the user device has been compromised and the random matrix has been exposed. We continue the training of the attack model until the average Euclidean distance between the attack profile and the corresponding artificially generated profile reached below a certain limit. Then, using the trained attack model, we recovered the plain profiles of the projected profiles of Group 1, and estimated the feature distribution similarity along with ground truth profiles using statistical test. We found that in both cases, only around 3.0%-12.0% of the features of a group of profiles in all three data sets are recoverable by this privacy attack.

Table 1: List of notations

| Notation | Meaning |
|---|---|
| $\mathbf{X}$ | A BA profile |
| $\mathbf{Y}$ | A verification data (profile) |
| $\mathbf{x_i}, \mathbf{y_i}$ | Data sample (vector) |
| $d$ | Dimension of a vector (total features) |
| $n, m$ | Number of vectors in a (verification) profile |
| $\mathbf{R}$ | Random matrix of dimension $k \times d$ |
| $\mathbf{x'_i}, \mathbf{y'_i}$ | Projected vector |
| $\mathbf{X'}, \mathbf{Y'}$ | Projected profile |
| $\mathsf{Ver}(\cdot, \cdot)$ | Verification algorithm |
| $\mathsf{C}(\cdot)$ | Neural Network classifier |
| $\hat{y}$ | Prediction vector |
| $\mathsf{M_P}(\cdot)$ | ML model for privacy attack |

**Notation.** Table 1 summarizes the notations used in this paper.

**Paper organization.** Section 2 describes the preliminary and related works. The privacy-preserving BA system is covered in Section 3 while the privacy attack is described in Section 4. Section 5 gives details of the experimental results. Finally, Section 6 concludes the paper.

## 2 PRELIMINARIES AND RELATED WORKS

The contexts and terminologies that are relevant to and frequently utilized in this paper are explained in this section.

### 2.1 BA System

A BA system collects a set of $N$ profiles during the registration phase and utilizes them to train an NN-based BA classifier $\mathsf{C}(\cdot)$. For a verification claim, a verification profile (data) is collected during the verification phase. A verification algorithm $\mathsf{Ver}(\cdot, \cdot)$ of the BA system uses the trained classifier $\mathsf{C}(\cdot)$ to generate the prediction vectors and leverage them for authentication decisions. The structure of a BA profile and an architecture of the BA system are shown in Figure 1 where the verifier is a server that performs authentication using a verification algorithm.

DEFINITION 2.1. *(Profile and verification data) A profile $\mathbf{X}$ in a BA system consists of a set of $m$ $d$-dimensional vectors $\mathbf{x_i}, i = 1, \cdots, m$, over $\mathbb{R}^d$, where each component of $\mathbf{x_i} = (x_{i,1}, x_{i,2}, \cdots, x_{i,d})$ represents the measurement of a feature value. The set $\{x_{i,j}, i = 1, \cdots, m\}$ corresponds to the samples of a feature $F_j$. So, the profile $\mathbf{X}$ can also be seen as a sequence of feature variables $\mathbf{X} = (F_1, F_2, \cdots, F_d)$ represented by their corresponding sample sets. A verification request generates a new verification profile $\mathbf{Y} = \{\mathbf{y_1}, \mathbf{y_2}, \cdots, \mathbf{y_n}\}$, a second measurement of behavioral data from the user $u$.*

A BA system either keeps all of the $N$ profiles in a profile database or uses them to train a classifier $\mathsf{C}(\cdot)$. A classifier-based BA system is useful for continuous authentication and significantly improves the system's performance.

DEFINITION 2.2. *(Verification algorithm) When a verification algorithm in a BA system receives a verification request $(u, \mathbf{Y})$, it uses the profile $\mathbf{X}$ of user $u$ and the presented verification data $\mathbf{Y}$ to determine the closeness between them. The result is presented as a score. If the verification algorithm utilizes the classifier $\mathsf{C}(\cdot)$ for verification, the output will be $n$ prediction vectors $\hat{y}$ for each $\mathbf{y_i} \in \mathbf{Y}$. The probability values in a prediction vector can also be considered as score. The final output of the verification algorithm is either to accept or reject the claim.*

A BA system must guarantee correctness and security properties.

(1) $\delta_1$-correctness: The probability that the classifier $\mathsf{C}(\cdot)$ of the BA verification algorithm incorrectly classifies $\mathbf{y_i} \in \mathbf{Y}$ for a valid verification request $(u, \mathbf{Y})$ which lead to the final verification decision as 0 (reject),

$$Pr[\mathsf{Ver}(\mathbf{Y}) = 0 \mid \mathsf{C}(\cdot) \text{ classify } \mathbf{y_i} \in \mathbf{Y} \text{ incorrectly }] \leq \delta_1. \quad (1)$$

The value of $\delta_1$ related to the correctness of $\mathsf{C}(\cdot)$ for all valid claims.

(2) $\delta_2$-security: The probability that the classifier $\mathsf{C}(\cdot)$ of the BA verification algorithm incorrectly classifies $\mathbf{y_i} \in \mathbf{Y}$ for an invalid verification request $(u, \mathbf{Y})$ which lead to the final verification decision as 1 (accept),

$$Pr[\mathsf{Ver}(\mathbf{Y}) = 1 \mid \mathsf{C}(\cdot) \text{ classify } \mathbf{y_i} \in \mathbf{Y} \text{ incorrectly }] \leq \delta_2. \quad (2)$$

The value of $\delta_2$ related to the correctness of $\mathsf{C}(\cdot)$ for all invalid claims.

In both cases, the probability is over all users and their valid and invalid verification requests.

*Measuring performance.* For a BA system, the above correctness and security errors are calculated using the False Rejection Rate (*FRR*) and False Acceptance Rate (*FAR*), respectively. $\delta_{FRR} = \delta_1$ will be calculated by the accuracy of the classifier $\mathsf{C}(\cdot)$ for all valid claims, and $\delta_{FAR} = \delta_2$ will be calculated by the accuracy of $\mathsf{C}(\cdot)$ for all invalid claims.

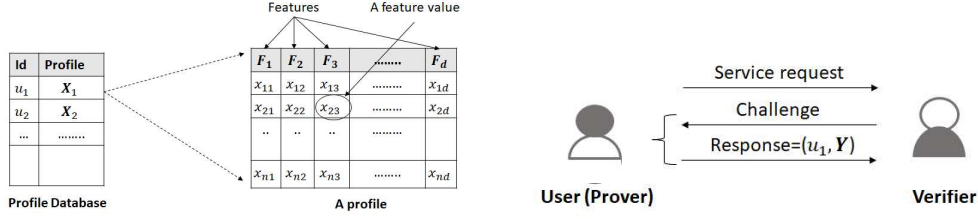

Figure 1: (A) A profile in a BA system, where each row represents a feature vector and each column represents the samples of a feature. (b) A verification algorithm generates scores based on the users' respond.

## 2.2 Machine Learning (ML) Classifier

A machine learning classifier $C(\cdot)$ learns the mapping between a set of *training data* and a set of classes and represents the knowledge by the parameters (independent variables) of the parametric functions. In an NN classifier, the input is a vector of dimension $d$, and the output is a $N$ dimensional *prediction vector*. The prediction vector, which is represented by a vector of $N$ probability values, reflects the classifier's belief that the input belongs to one of the $N$ classes of the NN classifier. The *verification algorithm* of the BA systems utilizes the prediction results of the NN classifier to make an authentication decision.

DEFINITION 2.3. *(NN classifier) A NN classifier $C(\cdot)$ uses a hierarchical composition of n parametric functions $f_i$ to model an input vector $\mathbf{x}$ where each $f_i$ is modeled using a layer of neurons and parameterized by a weight vector $\mathbf{w}_i$. The last layer of $C$ uses a softmax function $\sigma(\cdot)$ (also called a normalized exponential function) to encode the belief of the input being in each of the classes of the NN classifier and represented by the probability values of a prediction vector $\hat{\mathbf{y}}$.*

$$\hat{\mathbf{y}} = C(\mathbf{x}) = \sigma(f_n(\mathbf{w}_n, f_{n-1}(\mathbf{w}_{n-1}, \cdots f_2(\mathbf{w}_2, f_1(\mathbf{w}_1, \mathbf{x}))))) \quad (3)$$

A neuron is an elementary computing unit that uses a set of inputs, a set of weights, and an activation function. It translates the inputs into a single output, which can then be used as input for another layer of neurons. While the details can vary between neural networks, the function $f_i(\mathbf{w}_i, \mathbf{x})$ is always a weighted function in the form of $\mathbf{w}_i \mathbf{x}$. The weights for each neuron are adjusted throughout the training phase so that the final output of the network, the prediction vector $\hat{\mathbf{y}}$, is biased toward the value of the ground truth vector $\mathbf{y}$. The non-linear behavior in a neural network is accomplished by using an activation function (often a sigmoid function), to which the output of $f_i$ is passed and modified. This non-linear behavior allows neural networks to describe more complicated systems while still combining inputs in a simple fashion.

In a NN classifier $C(\cdot)$, the dimension of $\hat{\mathbf{y}}$ is equal to the dimension of $\mathbf{y} \in Y$. If $\mathbf{x}$ is from user $u_i$, then ideally the corresponding $\mathbf{y}$ will hold a probability value close to $\mathbf{y}[i] = 1.0$ and close to 0.0 in the rest of the classes. A loss function measures the (distribution) difference between $\mathbf{y}$ and $\hat{\mathbf{y}}$.

## 2.3 Random Projection

Random Projection (RP) is a distance-preserving transformation that projects vectors in a high-dimensional space to a lower-dimensional space using a random projection matrix.

DEFINITION 2.4. *RP uses a random matrix $\mathbf{R}^{k \times d}$, where $k < d$, to project a vector $\mathbf{x} \in \mathbb{R}^d$ to a new vector $\mathbf{x}' \in \mathbb{R}^k$ as*

$$\mathbf{x}' = \frac{1}{\sqrt{k}\sigma_r} \mathbf{R}\mathbf{x}, \quad (4)$$

where $\sigma_r$ is the standard deviation of the entries of a random matrix $\mathbf{R}$. The RP of a plain profile $\mathbf{X}$ can be represented as $\mathbf{X}' = \mathbf{R}\mathbf{X}$.

The important property of RP is that it preserves pair-wise Euclidean distances between the points in the metric space $\mathbb{R}^d$, up to an error that can be estimated for the dimension reduction parameter. The existence of distance-preserving dimension reduction transformations follows the following lemma:

**Johnson-Lindenstrauss(JL) Lemma [10].** Let $\epsilon \in (0, 1)$ and $\mathcal{M}_d \subset \mathbb{R}^d$ be a set of $n$ vectors and $k = \frac{4 \ln(n)}{\epsilon^2/2 - \epsilon^3/3}$. There exists a Lipshcitz mapping $f : \mathbb{R}^d \to \mathbb{R}^k$ such that for all $\mathbf{x}_i, \mathbf{x}_j \in \mathcal{M}_d$, the distance function $D$

$$(1 - \epsilon)D^2(\mathbf{x}_i, \mathbf{x}_j) \leq D^2(f(\mathbf{x}_i), f(\mathbf{x}_j)) \leq (1 + \epsilon)D^2(\mathbf{x}_i, \mathbf{x}_j).$$

The proof of the JL Lemma constructs the RP transform using matrices whose entries are sampled from the normal distribution $N(0, 1)$ [10].

## 2.4 Privacy-Preserving ML (PPML)

A machine-learning task often involves three different parties. They are the data owner or contributor, the calculation party, and the result query party. If all three roles are assumed to be played by the same entity, ML privacy is naturally preserved. However, these roles are split across two or more entities in real-world applications. In a BA system, the verifier serves as the computation party, and the input parties are the users who register in the system using their profiles. The query party can be either a user or an attacker. In all cases, the ML should ensure training and query data privacy. There are many works that take the issue of PPML into consideration and employ different techniques, such as differential privacy [13], homomorphic encryption [14], and multi-party computation [25] to ensure data privacy in ML model.

In a BA system, users are the owners of the data and the verifier is the computation party (verifier); therefore, the data must be protected during transit, idle, and in use phases. Some ML algorithms used by the computation party (BA verification algorithm), such as SVM and kNN, usually store feature vectors (or extracted features) inside the model, which puts the data at great risk. In those cases, an adversary can easily reconstruct the raw data from the stored features. The neural network used by the computation party does not store the feature vectors explicitly. However, a white-box access to the model helps the attacker reconstruct the model to recover the training data, which is called a model reconstruction attack. The results party should be restricted to black-box access to the model [16], in order to fend off such reconstruction attempts. Another option to restrict access to the model output is to combine $n$ results into a single output and then publish it [2]. The membership inference attack infers whether a sample was in the training set based on the ML model output and is out of the scope of this research. Differential privacy [13] is a popular approach to defending against membership inference attacks.

Reconstruction attacks construct the training data by using the knowledge of the features' vectors that are stored implicitly or explicitly inside the model. PPML techniques also allow multiple input parties to collaboratively train ML models without releasing their private data in its original form, which is called federated learning [5, 25]. Training data encryption, such as homomorphic encryption [14, 24] is also a popular approach to keeping the training data secure. Some other PPML techniques are also proposed in [4][7].

# 3 PRIVACY PRESERVING BA SYSTEM

## 3.1 Requirements of Proposed Scheme

We consider the data privacy protection scenario inside the BA systems, which aims to fulfill the following requirements:

(1) **Usability:** In a privacy-preserving BA system, a rightful user with a correct secret random matrix $\mathbf{R}$ will not experience any performance (correctness) degradation during verification. The key generation, key management, and random projection should also be simple and easily manageable at the user end.
(2) **Unusability:** Ideally, for a profile with the wrong random matrix, the performance of the model should be heavily dropped, which directly links with the security of the privacy-preserving BA system.
(3) **Changeability.** The ability to refresh a user profile while maintaining the correctness and security properties of the system. For RP-based NN classifier, the changeability property of the system can be defined as

DEFINITION 3.1. ($\delta_3, \delta_4$-changeability) A BA system provides $\delta_3$-changeability if it satisfies the following:
The probability that the updated classifier $C(\cdot)$ of the BA verification algorithm incorrectly classifies $Y'_1 = R_1Y$ to 1 (accept) and $Y'_2 = R_2Y$ to 0 (reject) when $X'_1 = R_1X$ is used to train $C(\cdot)$ and later $X'_2 = R_2X$ to update $C(\cdot)$.

$$Pr[\text{Ver}(R_1Y) = 1] \leq \delta_3 \ \& \ Pr[\text{Ver}(R_2Y) = 0] \leq \delta_4$$

The FRR and FAR will mostly determine the usability and unsuitability properties of the privacy-preserving BA system. A system that has lower FRR and FAR is more usable for valid users and less usable for invalid users. The changeability property can also be evaluated by the FRR and FAR for those claims where a valid user uses a correct and incorrect random matrix to project the verification data.

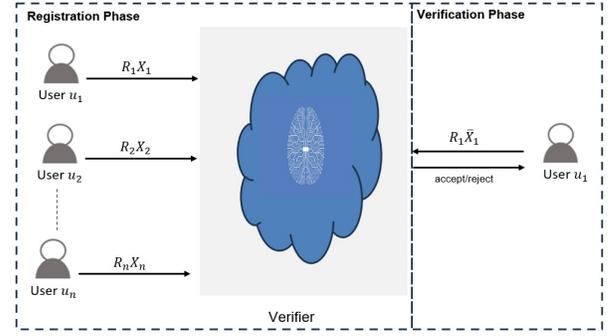

**Figure 2: Privacy-preserving BA system. During the registration phase, the verifier will use projected profiles to train a classifier. The verifier will output an accept or reject decision for the verification claim on the projected verification data.**

## 3.2 System Overview

An overview of the privacy-preserving BA system is depicted in Figure 2. In the proposed system, during the registration phase, users will transform their profiles using a secret random matrix and send the projected profiles to the verifier. For such $N$ projected profiles, an NN-based classifier $C(\cdot)$ will be trained. During the verification phase, the verification profile will also be transformed by the same random matrix that was used in the registration phase. The verifier will output an accept or reject decision based on the outputs of the classifier $C(\cdot)$.

The registration phase of the privacy-preserving BA system consists of three main tasks:

(1) **Random matrix generation:** Each user will utilize their devices to create a secret, $k \times d$-dimension random matrix. The users will safely store the random matrix on their devices for further usage. For faster computation in low-end devices, a discretized form of $N(0,1)$ will be used to generate the random matrix $\mathbf{R}$ (see Equation 5).

$$Pr(x = +1) = \frac{1}{2\phi}, \ Pr(x = +0) = 1 - \frac{1}{\phi}, \ Pr(x = -1) = \frac{1}{2\phi} \quad (5)$$

To ensure the distance preserving property, we follow the Theorem 3.1.

THEOREM 3.1. ([1]) Suppose $\mathcal{M}_d \subset \mathbb{R}^d$ be a set of $n$ vectors projected onto $\mathbb{R}^k$ using the transform $f : \mathbb{R}^d \to \mathbb{R}^k$ defined as $f(\mathbf{x}') = \frac{1}{\sqrt{k}\sigma_r}\mathbf{R}\mathbf{x}$ for $\mathbf{x} \in \mathcal{M}_d$, where $\mathbf{R}$ is a $k \times d$ matrix generated using the distribution given in equation (1) and $\sigma_r$ is the standard deviation of entries of R. Given $\epsilon, \beta > 0$ let $k_0 = \frac{4+2\beta}{\epsilon^2/2-\epsilon^3/3}\log(n)$. If $k \geq k_0$ then with probability at least $1 - n^{-\beta}$ for all $\mathbf{x}_i, \mathbf{x}_j \in \mathcal{M}_d$ we will have:

$$(1-\epsilon)D^2(\mathbf{x}_i, \mathbf{x}_j) \leq D^2(f(\mathbf{x}_i), f(\mathbf{x}_j)) \leq (1+\epsilon)D^2(\mathbf{x}_i, \mathbf{x}_j)$$

For $\phi = 3$, the distance-preserving property of the resulting RP is shown in [1]. We will also keep the value of $\phi$ same in our experiments.

(2) **Random projection:** User $u_i$ will then project the profile $\mathbf{X}_i$ by the randomly generated matrix $\mathbf{R}_i$ as $\mathbf{X}'_i = \mathbf{R}_i\mathbf{X}_i$. RP will follow the Lipshcitz mapping $f : \mathbb{R}^d \to \mathbb{R}^k$ and will map the profile from $d$ to $k$ dimension. The value of $k$ depends on the number of vectors in the profile (see Johnson-Lindenstrauss (JL) Lemma). In [26], authors show that applying RP in the BA system as $\mathbf{X}' = \mathbf{RX}$ maintains the correctness property of the system. After projection, the user will send the project data to the verifier.

(3) **Train an NN classifier.** The verifier will gather all $N$ projected profiles and use them to train a NN-based classifier $\mathsf{C}(\cdot)$. Here, a verifier can be a third-party service provider who can deploy the trained classifier as MLaaS. In this case, the user can perform queries via APIs for authentication without the need to implement the entire framework.

At the verification phase, user $u_i$ will transform the verification profile as $\mathbf{Y}'_i = \mathbf{R}_i \mathbf{Y}_i$ before sending the verification request $(u_i, \mathbf{Y}'_i)$ to the verifier. The verifier will return an accept or reject decision based on the outputs of the classifier $\mathsf{C}(\cdot)$ for the input $\mathbf{Y}'_i$. In this case, the verifier will be given $n$ prediction vectors for $n$ vectors of $\mathbf{Y}'_i$. The verifier will then aggregate the classification decisions into a single binary decision (0 for reject and 1 for accept) and return it to the user.

To refresh the trained classifier, user $u_i$ will just change the random matrix to $\mathbf{R}_j$ which will produce a new projected profile as $\mathbf{Y}_j = \mathbf{R}_j \mathbf{X}$. The procedure is the same for all other users in the system. The new projected profiles will then be used to update the NN classifier $\mathsf{C}(\cdot)$. However, the correctness and security properties of the BA system should stay almost the same even for updated projected profiles.

## 3.3 Robustness Against Attack

**Key Estimation Attack.** We consider a scenario where attacker uses brute-force attacks to recover the secret key. Here the dimension of the random matrix $\mathbf{R}$ is $k \times d$, and each component of the matrix is either 0 or 1. So, the worst-case complexity to guess the random matrix is $O(k \times d)$. An attacker will need to guess 140 bits, which is more than the 128-bit key of AES, if $d = 14$ and $k = 10$. In our case, the size of the $d$ and $k$ is larger than the given example. Moreover, key estimation attacks do not guarantee that the attacker will find the correct key because the attacker does not know the actual performance of the correct key. We also show that a privacy attacker will not able to recover plain profile from projected profile event if $\mathbf{R}$ is known.

**Reconstruct Profile by Minimum-norm Solution.** Let $\mathbf{X}' = \mathbf{RX}$, where $\mathbf{X}' \in \mathbb{R}^d$ and $R \in \mathbf{R}^{k \times d}$, $k < d$, and $\mathbf{X}'$ is known. If $\mathbf{X}$ is not known, this system of linear equations has d-k degrees of freedom. Among all solutions, the solution $\hat{\mathbf{X}} = R^T(RR^T)^{-1}\mathbf{X}'$, known as the *minimum-norm solution* is the best known solution that minimizes the Euclidean norm of the solution $\|\hat{\mathbf{X}}\| = \sqrt{\sum_{t=1}^{m} \mathbf{X}_i'^2}$ [11] to recover $\mathbf{X}$. However, by deriving theoretical bounds on privacy and also by verifying the results experimentally, the authors in [30] shown that for behavioral data, RP is an effective privacy-preserving transformation and minimum-norm solution based attack can not recover the plain profile from projected profile.

## 4 ML-BASED PRIVACY ATTACK

### 4.1 Attacker's knowledge.

We assume that the privacy attacker has the following knowledge and capabilities to recover the plain profile from the projected profile:

(1) The attacker knows the BA system and the algorithm used during the verification phase. The attacker is also aware of the architecture and input-output structure of the NN classifier.
(2) Though the attacker does not have access to plain profiles, attacker can have access to projected profiles of the BA system. The attacker can get the projected profiles from the honest-but-curious verifier or from a model inversion attack. Here, the honest-but-curious verifier will follow the protocol but would like to glean information about users from their projected data.
(3) Typically, the BA system's profile generator is accessible to the general public. A profile generator is software that is used by the BA system to collect behavioral data from users. The attacker will outsource the profile generator to collect attack data from the users whose profiles are not used to train the classifier.
(4) The attacker is aware of the distribution and dimension of $\mathbf{R}$, as this is public information. The attacker will use this information to generated random matrix. In the worst case, the attacker also knows secret $\mathbf{R}$. The attacker will use $\mathbf{R}$ to obtain the projected version of the attack data.

### 4.2 ML Model for Privacy Attack

The goal of this privacy attack is to recover plain profiles from the projected profiles of the BA systems. For this privacy attack, we train an ML model called $\mathsf{M_P}(\cdot)$. The input of the model is a projected vector, and the output is plain vector. The model will be trained on the projected version of the attack data (profiles) to recover their plain version to know the behavioral pattern of the users. During training phase, the attack data (profiles) and their projected version will transfer knowledge to the attack model about the inversion of the RP data. Details about each step are given below:

**Step1: Generating attack data.** Each BA system has a profile generator algorithm that collects the user's behavioral data and constructs the profile. Profile generators are usually mobile applications (software), and they are publicly available. The attacker can use the software to collect the required attack data using any out-souring platform, such as Amazon Mechanical Turk. There is no limitation on the number of attack profiles, though more attack data means a more generalized attack model.

**Step2: RP on attack data.** For each attack profile, the attacker will either generate a random matrix or use the users' secret matrix. The attacker will use those random matrix to project the attack profiles. An attacker may employ more than one random matrix for an attack profile in order to have a larger projected profile of their attack. This will also increase the training data for the attack model.

**Step3: Train the attack model.** The attacker will use the projected version of the attack profiles to train the attack model. Figure 3

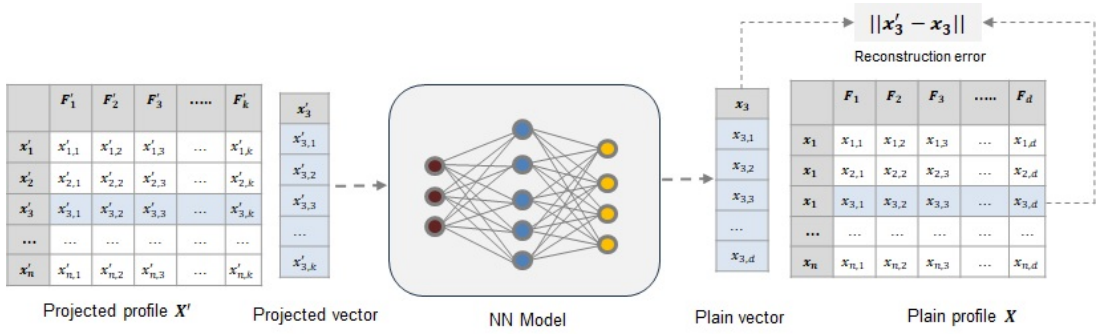

Figure 3: For a projected attack profile X′ with $m$ projected vector, $M_P(\cdot)$ will generate $m$ plain vectors. The error in the prediction will then be measured by measuring the reconstruction error of the plain vector of the attack profile.

shows the training process of $M_P(\cdot)$. The goal of the attacker is to generate plain vectors from projected vectors of a projected profile so that a new profile generated on those plain vectors will be close to the ground truth profile. Here, the ground truth profile is an attack profile. The training process of $M_P(\cdot)$ will continue till we get very close profiles to the ground truth.

**Step3: Generate plain profile.** We will use the train attack model $M_P(\cdot)$ to generate the plain profiles from the projected profiles that were used in the BA system to train the NN classifier. The success of this attack depends on the closeness of the new generated profiles with the original plain profiles of the users.

### 4.3 Privacy Evaluation

The privacy evaluation will be based on the closeness of the profiles. Our privacy-preserving BA system will ensure the privacy of the training data (profiles) if the recovered profiles are not close to the users' plain profiles. Here, closeness will be measured by the statistical closeness between the distribution of the features of the recovered profile and the plain profile. In [31], the authors named this type of privacy as *Distribution-privacy*, where the goal of the attacker is to recover the distribution of features of a plain profile from its projected profile.

DEFINITION 4.1. *($\epsilon$-Distribution-privacy:) Let's say we apply a privacy attack model $M_P(\cdot)$ to a multivector profile X′ to produce X. Then $M_P(\cdot)$ provides $\epsilon$-Distribution-privacy against an attacker if the best ML-based strategy of the distribution attacker results in an estimated profile in which at most $\epsilon$-percent of features pass a statistical closeness test with the corresponding features in the original profile.*

As the BA profile is a *multivector profile*, the goal of the adversary will be to learn the distribution of the features. Definition 4.1 introduces the notion of $\epsilon$-Distribution-Privacy, where $\epsilon$ is the fraction of features (in the feature vector) that remain "close" to their original distributions, given the adversary's knowledge. We can use any statistical hypothesis test to estimate the closeness between two features. We consider two types of *adversary knowledge*: (i) the adversary knows the distribution of random matrices R but does not know the R that is assigned to the user, and (ii) the adversary knows R. This second scenario corresponds to the most extreme scenario, in which the user device has been compromised and R has been exposed. In the case of unknown R, the attacker will use their own generated random matrix to project the attack profiles and then use them to train the attack model. On the other hand, if the adversary knows the users' secret random matrix R, adversary will use those matrices to project the attack profiles and then train the attack model. In both situations, a privacy-preserving BA system will prevent a privacy attack from recovering more features from each profile than *epsilon* on average. The second attack model is expected to have better performance than the first model.

## 5 EXPERIMENTAL RESULTS

We have implemented and evaluated our proposed approach to behavioral biometric data. We get swipe and voice pattern data from [18] and drawing pattern data from [20]. The swipe and voice data sets have 10,320 observations (vectors) from 86 users, with 120 observations per user. Drawing data has 80–240 observations per user, and there are 193 distinct users. Swipe data contains 33 features extracted from swipe gestures, and voice data contains 104 voice features. On the other hand, drawing pattern data has 65 features captured from users' responses in a circle drawing game.

**Experiment setup.** We downloaded and cleaned the data [1] before using them. To reduce the effect of biases that are the result of features having different ranges, we also normalized the feature values so that all feature ranges coincide with [0,1]. From each profile, we separated 20.0% of the data for testing purposes.

*Data oversampling.* For NN experiments, we need sufficient data for each profile. We used the Synthetic Minority Over-sampling Technique (SMOTE) [8] in each profile to increase the sample size.

---
[1]We replaced "NaN" and "Infinity" with zero and dropped duplicate row

| Data Set | $k_0$ | $n$ | $\epsilon$ | $\beta$ | $1 - n^{-\beta}$ |
|---|---|---|---|---|---|
| Swipe Pattern | 30.0 | 300 | 1.0 | 0.5 | 0.94 |
| Voice Pattern | 73.0 | 200 | 0.5 | 1.0 | 0.99 |
| Drawing Pattern | 46.0 | 300 | 0.7 | 1.0 | 0.99 |

Table 2: Minimum acceptable value of $k$ in RP for different data sets.

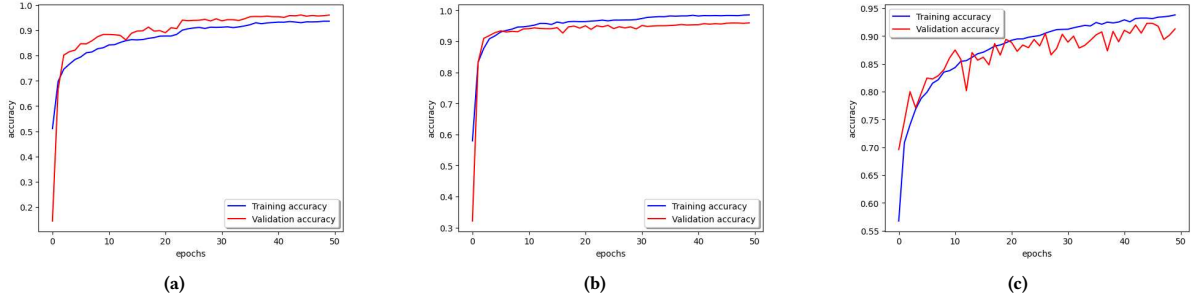

Figure 4: Training and validation accuracy of the (a) swipe, (b) voice, and (c) drawing data classifiers. In 50 epochs of training, all classifiers can achieve 93.58%, 98.53%, and 93.77% of training accuracy and 96.0%, 95.96%, and 91.25% of validation accuracy, respectively. Test accuracy follows almost same trend.

SMOTE is an oversampling algorithm that produces new data that lies between any two existing data samples of a profile. So, SMOTE does not add any new knowledge to a profile. In our experiments, we made sure that each training profile in swipe and drawing data sets had a minimum of 300 data samples and 200 data samples in voice data profile [2]. We also verify the accuracy and security of both BA systems after applying SMOTE.

*Training data and attack data.* We divided all profiles in each data set into two groups: (i) *Group 1-* all profiles and their projected version of this group are used to train and validate the NN classifier of the BA system, and (ii) *Group 2-* all profiles and their projected version of this group are used as attack data. In the first group, there are around 80.0% of the profiles (68 swipe and voice profiles, and 155 drawing pattern profiles), and the rest of the other 20.0% profiles are in Group 2 (18 swipe and voice profiles, and 38 drawing pattern profiles).

*Random matrix.* For profile projection, we generated random matrices by using discretized from of $N(0, 1)$. Each profile (both training and attack profiles) of all three data sets uses a distinct random matrices **R** for RP. We used the Johnson-Lindenstrauss (JL) Lemma to calculate the minimum value of $k$ for a $k \times d$-dimension random matrix. Table 2 shows the minimum value of $k$, i.e., the value of $k_0$ and they are 30.0, 73.0, and 46.0, respectively for swipe, voice, and drawing pattern-based profiles that ensure a minimum 0.94, 0.99, and 0.99 distance-preserving probability in RP. For our experiment, we chose $k$ as 30, 94, and 56 for swipe, voice, and drawing pattern data.

## 5.1 Design and Train Classifier

In this section, we design the NN architectures for the BA classifiers. We then train and validate them by using the plain data samples from Group 1. We also test the performance of the classifiers by using the test data. The objective of this experiment is to demonstrate that data oversampling has no abrupt impact on the system's performance, and that the system can continue to operate with almost the same performance as described in the original study.

---

[2]For less features in a data set, it requires more data samples in a profile to train a NN classifier

**Experiment 1: Classifier Design.** To validate the effectiveness of SMOTE and to check the correctness and security properties of those systems on plain data sets, we trained a NN-based BA classifiers. For all three data sets, we designed three separate hierarchical NN architectures. The concept behind constructing a hierarchical model is that each higher-level layer in a NN classifier captures more intricate non-linear properties from training data. All three architectures have multiple dense, batch-normalization, and activation (ReLU) layers. The dense layer offers learning features from all combinations of the features of the previous layer. The ReLU layer works as an activation function of the soft boundary, and the batch-normalization layer normalizes the data. In order to address the issue of over-fitting, we added dropout layers in the network. All classifiers have five stacks of layers, and each layer has 128, 256, 512, 256, and 128 nodes, respectively. The drop-out value for all layers is 0.1. A *softmax function* layer is used in the last layer of each classifier to represent the output probability distribution in the prediction vector. Figure 8 in Appendix A shows the NN architectures of a BA classifiers.

**Experiment 2: Train classifier and estimate accuracy.** In all three classifiers, we use Group 1 data to train and validate the accuracy of the classifiers. From the data samples of each profile in Group 1, we use 80.0% of the data for training and 20.0% for validation purposes. All three classifiers (swipe, voice, and drawing pattern) achieved 93.58%, 98.53%, and 93.77% classification accuracy, and 96.0%, 95.96%, and 91.25% validation accuracy, respectively. Figure 4) shows the trend of training and verification accuracy in 50 epochs of training for all three classifier. For the test data that we separated earlier, the classification accuracy is 91.69%, 98.10%, and 93.93%, respectively, for swipe, voice, and drawing pattern data. According to our definition, this will be equivalent 8.31%, 1.90% and 6.07% FRR of the BA system. For swipe and voice data, the authors of [17] reported around 3.52%-5.0% of FRR, and [21] reported 4.69% FRR for drawing pattern data.

To ensure that none of the profiles of Group 2 are used to train their corresponding classifier, we estimate the classification accuracy of all three classifiers by using Group 2 profiles. In this case, the label of Group 2 data was randomly replaced by the label of Group 1 for an invalid verification claim. Experimental results show that swipe, voice, and drawing pattern classifiers have only 0.90%, 0.94%, and 0.69% classification accuracy, which is equivalent to 0.90%, 0.94% and 0.69% FAR of the BA system. For the attack data, this result is what we expect to get.

| Data Set | Correctness | Security |
|---|---|---|
| Swipe Pattern | 1.07% FRR | 0.23% FAR |
| Voice Pattern | 0.44% FRR | 0.45% FAR |
| Drawing Pattern | 1.03% FRR | 0.33% FAR |

Table 3: Correctness and security of the privacy-preserving BA system. All three systems based on three different data sets and they have very low FAR and FRR.

## 5.2 Design & Train Privacy-preserving Classifier

In this section, we design the NN architecture for privacy-preserving BA systems using projected profiles of Group 1. We then train and validate the performance of the classifier on projected data. We used both correct and incorrect random matrices for test data projection and then tested the correctness and security of the system. In this section, we also examine the system's changeability property.

**Experiment 1: Design and Train a privacy-preserving classifier.** We create three distinct hierarchical NN architectures for three projected data sets. The NN layers used in all three designs are almost the same as those NN layers used for plain data: dense, batch-normalization, activation (ReLU), and dropout layers. We just made a few adjustments in the number of stacks of layers in NN and also in the number of nodes in each layer. We also updated the dropout rate in a few architectures. A *softmax function* layer is also used as the last layer of each NN. Figure 9 in Appendix B shows an NN architecture for privacy-preserving BA classifiers.

For all three classifiers, we use the projected version of Group 1 data to train and validate the accuracy of the classifiers. From the data samples of each profile in Group 1, we use 80.0% of the projected data for training and 20.0% for validation purposes. For the correctness test, we projected the test profiles by using the same random matrices that were used in the training phase. On the other hand, we projected the test profiles using a different set of random matrices to test the security of the system. For training data, all three classifiers (swipe, voice, and drawing pattern data) achieve 96.42%, 95.65.0%, and 97.34% classification accuracy, and 98.95.0%, 95.10%, and 96.56.0% validation accuracy, respectively. In RP, each user uses their own matrix, which has somewhat increased the distances among the profiles and brought system accuracy above 95.0%. For the test data projected by the correct random matrix, the classification accuracy is also 98.93%, 99.56%, and 98.97%, respectively, which implies that the system has an FRR below 2.0%. On the other hand, system accuracy reduced to 0.23%, 0.45% and 0.33%, respectively in all three cases when we used separate set of random matrices in the projection, which is equivalent to 0.23%, 0.45% and 0.33% FAR, respectively. Table 3 summarizes the correctness and security properties of the privacy-preserving BA system. Even when valid users used the incorrect random matrix to project their legitimate profiles, the accuracy of these three systems reduced to 1.3%, 0.5%, and 0.7% respectively, which is expected. Which also protects the users from an attack when their profiles are leaked unintentionally.

**Experiment 2: Model Update** We evaluate the changeability property of the privacy-preserving BA system for all three data sets. For our experiments, we employed a different random matrix than the one used previously to project the training profile. We then update the trained classifier with the new projected profiles. One can follow any of these two approaches to updating the classifier: (i) add one or more new layers to the existing trained classifier and then train those layers by transfer learning; or (ii) update all weights of the existing classifier. As it does not take too long to update a BA classifier, we followed the later approach. Within 20 epochs of the update all newly trained classifiers began to offer us very good training and validation accuracy. The training and validation accuracy of the new NN classifier is 90.22% and 97.93% for swipe, 90.32% and 93.45% for voice, and 88.60% and 90.56% for drawing data. We also test the model security using the previously projected test data that gives us close to 0.6% FAR for swipe, and 0.12% FAR for voice, and 3.6% FAR for drawing data set which is expected. However, if we project the test data by the new random matrices, the correctness of the system increases to 95.99% for swipe, 98.67% for voice and 97.92% drawing pattern data which is equivalent to 4.01%, 1.33% and 2.08% FRR, respectively. These two new correctness (FRR) and security properties (FAR) validate the changeability property of privacy-preserving BA systems by being very similar to the previous estimated correctness and security properties.

## 5.3 Privacy Attack

For a privacy attack, we train an NN-based attack model using the projected version of the attack data and then use the projected training data of the BA classifier $C(\cdot)$ to recover the plain profiles to know the behavioral pattern of the users.

**Experiment 1: Train an Attack Model.** We create three distinct hierarchical NN architectures for privacy attacks using three different attack data sets. The idea behind building a hierarchical model is that each higher-level layer in an NN captures a more complex non-linear relation between the plain and their projected profiles. All three architectures have multiple dense, batch-normalization, activation (ReLU) layers. The input dimensions of swipe, voice, and drawing-based privacy attackers are 30, 94, and 56, respectively, and the output dimensions are 33, 104, and 65, respectively. All three classifiers have four stacks of layers, and each layer has 128, 256, 256, and 128 nodes, respectively. The last layer uses the "sigmoid" function to keep the output in [0,1]. Figure 10 in Appendix C shows an NN architectures of the inverse classifiers.

We used a projected version of attack data (profiles) to train the attack model. We considered two cases to project a profile: (i) case 1- adversary only knows the distribution of **R** and uses this information to generate a set of new **R**, and (ii) case 2- adversary knows users' secret $R$ and uses them to project the attack profiles [3]. The projected profiles are then used as an input to the NN-based attack model, and the corresponding plain profiles of the attack profile are used as a ground truth. Our goal is to train the NN model to generate the plain profiles from the projected profiles. We use "mean-square-error" as a loss function in all three cases and "RMSProp" as an optimizer with a learning rate of 0.001.

---

[3]In our experiment, the number of secret **R** is higher than the number of attack profiles. We use more than one **R** to project an attack profile.

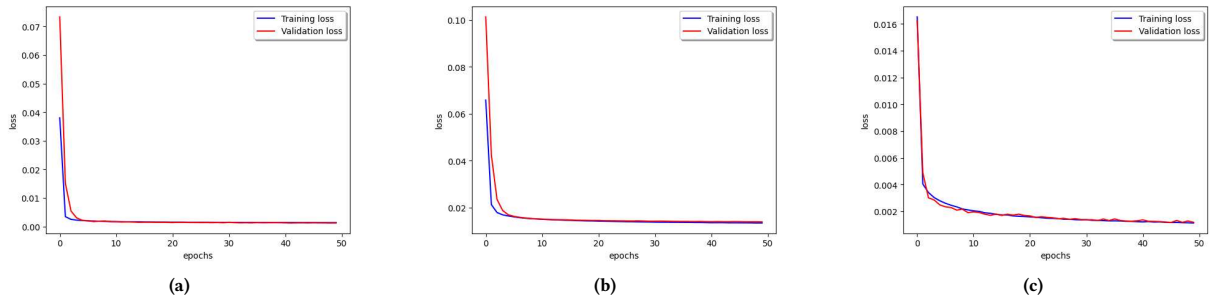

Figure 5: Training and validation loss of the attack model for (a) swipe, (b) voice, and (b) drawing pattern data. In 50 epochs of training, the training loss of all NN models was reduced to 0.0014, 0.013, and 0.0011, respectively and validation loss 0.0013, 0.013, and 0.0012, respectively.

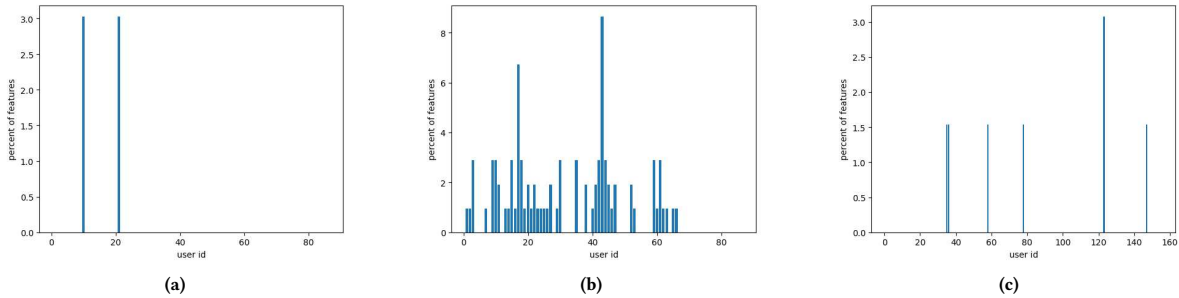

Figure 6: NN-based privacy attackers can recover features from projected profiles when the distribution of R is known. However, the percentage of recovered features is in the range 3.0% - 9.0% which is not sufficient to know details about the user's behavioral pattern.

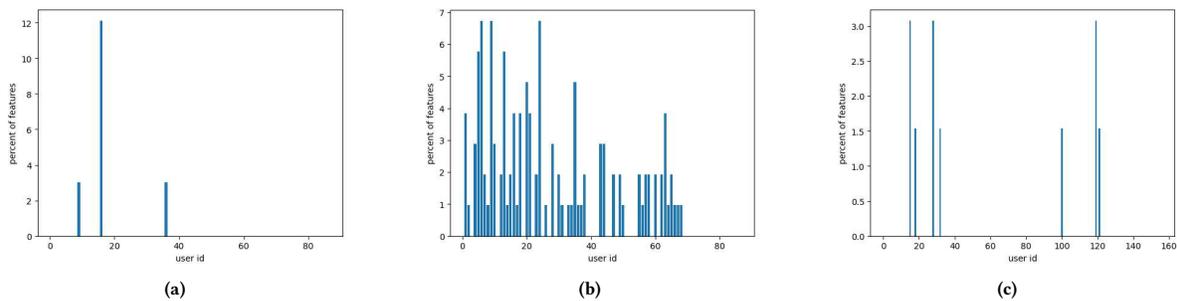

Figure 7: NN-based privacy attackers can recover features from more projected profiles when R is known. However, the percentage of recovered features is still low and is in the range 3.0% - 12.0% which is not sufficient to know details about the user's behavioral pattern.

For Case 1, after 50 epochs of training, the training loss of the NN-based attack model was reduced to 0.0014, 0.013, and 0.011, respectively, for the swipe, voice, and drawing-based attack data, while the validation losses became 0.0013, 0.013, and 0.0012, respectively. Figure 5 shows the graph for all three losses in 50 epochs of training. The value of the loss depends on how well the model is trained and the number of features in the data set. That is why we see the loss of voice data is bigger than other two data sets. For Case 2, we get almost the same training and validation losses as for in all three experiments, as we changed the random matrix, the attack data is same in both experiments.

**Experiment 2: Recover Plain Profile.** To recover plain profiles from the projected profiles of Group 1, we used the trained attack model. For each projected profile, the attack model will produce a plain profile. We then estimate the *Distribution-privacy* by comparing the feature distribution between recovered and ground truth profiles. To estimate the distribution similarity, we used the Kolmogorov-Smirnov test (KS test), a hypothesis-based closeness test. A single provability value (P-value) will be produced the KS-test for two input features where the first one will be drawn from the recovered profile, and the second one will be from the ground truth profile. If the P-value is less than 0.05 (not statistically significant), we rejected the null hypothesis that two features follow the same distribution. We repeated the test for all features in a profile and then count the percentage of features that passed the hypothesis test.

Figure 6 (a-c) shows the percentage of features that passed the null hypothesis test for all profiles in three data sets when the attacker only knows the distribution of **R**. In all three experiments, the NN-based privacy attack can only recover features from a few profiles. There are 2 swipe and 43 voice profiles out of 86 profiles, and 6 drawing profiles out of 155 profiles from which the attacker was able to recovered the features. However, the percentage of recovered features is only in the range 3.0% -9.0%.

Figure 7 (a-c) shows the percentage of features that passed the null hypothesis test for all profiles in three data sets when the attacker knows **R**. In these experiments, the NN-based privacy attack can recover features from a number of more profiles than case 1. There are 3 swipe and 48 voice profiles out of 86 profiles, 7 drawing profiles out of 155 profiles. However, the percentage of recovered features is still low and in the range 3.0% -12.0%. and the overall average is higher than the case 1. Our NN-based privacy-attack model reported almost similar success rate for the multivector profiles that were reported in [30]. However, they use a minimum-norm solution-based attack model, and in our case, it is an NN-based attack model. Moreover, in both of our attack scenarios, the features that were recovered are insufficient to know details about the user's behavioral pattern or to be used in a subsequent attack, like an impersonation attack.

## 6 CONCLUSION

We constructed and trained a privacy-preserving BA classifier. The outputs of this classifier were then used by the verification algorithm for authentication decisions. To protect the training and test data, users transformed the profiles before sending them to the verifier. Here the transformation was non-crypto-based, required less computation, and preserved the pair-wise distance among the profiles in another dimension. The proposed approach also protected the system from NN-based privacy attackers, where attackers know either secret random metrics or their distributions. In our experiments, we used swipe, voice, and drawing pattern data and achieved an accepted privacy and security guarantee. This strategy can be used to safeguard biometric features like face and fingerprint features in addition to the existing BA features. Our work presents several research issues, such as (i) which types of features are more vulnerable to this privacy attack and (ii) how to design a more effective privacy attack and defend the data against such attacks.

## REFERENCES

[1] Dimitris Achlioptas. 2003. Database-friendly random projections: Johnson-Lindenstrauss with binary coins. *J. Comput. System Sci.* 66, 4, 671–687.

[2] Mohammad Al-Rubaie and J Morris Chang. 2016. Reconstruction attacks against mobile-based continuous authentication systems in the cloud. *IEEE Transactions on Information Forensics and Security* 11, 12 (2016), 2648–2663.

[3] Mohammad Al-Rubaie and J Morris Chang. 2019. Privacy-preserving machine learning: Threats and solutions. *IEEE Security & Privacy* 17, 2 (2019), 49–58.

[4] Dan Bogdanov, Liina Kamm, Sven Laur, and Ville Sokk. 2018. Implementation and evaluation of an algorithm for cryptographically private principal component analysis on genomic data. *IEEE/ACM transactions on computational biology and bioinformatics* 15, 5 (2018), 1427–1432.

[5] Keith Bonawitz, Vladimir Ivanov, Ben Kreuter, Antonio Marcedone, H Brendan McMahan, Sarvar Patel, Daniel Ramage, Aaron Segal, and Karn Seth. 2017. Practical secure aggregation for privacy-preserving machine learning. In *proceedings of the 2017 ACM SIGSAC Conference on Computer and Communications Security*. 1175–1191.

[6] Joseph Bonneau et al. 2014. Privacy concerns of implicit secondary factors for web authentication. In *SOUPS Workshop on "Who are you?!": Adventures in Authentication*. Menlo Park, CA., 1–2.

[7] Raphael Bost, Raluca Ada Popa, Stephen Tu, and Shafi Goldwasser. 2014. Machine learning classification over encrypted data. *Cryptology ePrint Archive* (2014).

[8] Nitesh V Chawla, Kevin W Bowyer, Lawrence O Hall, and W Philip Kegelmeyer. 2002. SMOTE: synthetic minority over-sampling technique. *Journal of artificial intelligence research* 16 (2002), 321–357.

[9] Penny Chong, Yuval Elovici, and Alexander Binder. 2019. User authentication based on mouse dynamics using deep neural networks: A comprehensive study. *IEEE Transactions on Information Forensics and Security* 15 (2019), 1086–1101.

[10] Sanjoy Dasgupta and Anupam Gupta. 2003. An elementary proof of a theorem of Johnson and Lindenstrauss. *Random Structures & Algorithms* 22, 1 (2003), 60–65.

[11] James W Demmel and Nicholas J Higham. 1993. Improved error bounds for underdetermined system solvers. *SIAM J. Matrix Anal. Appl.* 14, 1 (1993), 1–14.

[12] Yunbin Deng and Yu Zhong. 2015. Keystroke dynamics advances for mobile devices using deep neural network. *Recent Advances in User Authentication Using Keystroke Dynamics Biometrics* 2 (2015), 59–70.

[13] Cynthia Dwork, Frank McSherry, Kobbi Nissim, and Adam Smith. 2006. Calibrating noise to sensitivity in private data analysis. In *Theory of Cryptography: Third Theory of Cryptography Conference, TCC 2006, New York, NY, USA, March 4-7, 2006. Proceedings 3*. Springer, 265–284.

[14] Zekeriya Erkin, Thijs Veugen, Tomas Toft, and Reginald L Lagendijk. 2012. Generating private recommendations efficiently using homomorphic encryption and data packing. *IEEE transactions on information forensics and security* 7, 3 (2012), 1053–1066.

[15] Mario Frank, Ralf Biedert, Eugene Ma, Ivan Martinovic, and Dawn Song. 2013. Touchalytics: On the applicability of touchscreen input as a behavioral biometric for continuous authentication. *IEEE Transactions on Information Forensics and Security* 8, 1 (2013), 136–148.

[16] Matt Fredrikson, Somesh Jha, and Thomas Ristenpart. 2015. Model inversion attacks that exploit confidence information and basic countermeasures. In *Proceedings of the 22nd ACM SIGSAC conference on computer and communications security*. 1322–1333.

[17] Sandeep Gupta, Attaullah Buriro, and Bruno Crispo. 2019. DriverAuth: A risk-based multi-modal biometric-based driver authentication scheme for ride-sharing platforms. *Computers & Security* 83 (2019), 122–139.

[18] Sandeep Gupta, Attaullah Buriro, and Bruno Crispo. 2020. A chimerical dataset combining physiological and behavioral biometric traits for reliable user authentication on smart devices and ecosystems. *Data in brief* 28 (2020), 104924.

[19] Md Morshedul Islam and Reihaneh Safavi-Naini. 2016. POSTER: A behavioural authentication system for mobile users. In *Proceedings of the 2016 ACM Conference on Computer and Communications Security (CCS '16)*. ACM, 1742–1744.

[20] Md Morshedul Islam and Reihaneh Safavi-Naini. 2020. Scalable Behavioral Authentication Systems. IEEE Access. manuscript submitted for review.

[21] Md Morshedul Islam and Reihaneh Safavi-Naini. 2021. Model Inversion for Impersonation in Behavioral Authentication Systems.

[22] Dawoon Jung, Mau Dung Nguyen, Jooin Han, Mina Park, Kwanhoon Lee, Seonggeun Yoo, Jinwook Kim, and Kyung-Ryoul Mun. 2019. Deep neural network-based gait classification using wearable inertial sensor data. In *2019 41st Annual International Conference of the IEEE Engineering in Medicine and Biology Society (EMBC)*. IEEE, 3624–3628.

[23] Weizhi Meng, Duncan S Wong, Steven Furnell, and Jianying Zhou. 2014. Surveying the development of biometric user authentication on mobile phones. *IEEE Communications Surveys & Tutorials* 17, 3 (2014), 1268–1293.

```
Layer (type)                 Output Shape              Param #
=================================================================
dense_50 (Dense)             (None, 128)               4352
batch_normalization_41 (Bat  (None, 128)               512
chNormalization)
activation_41 (Activation)   (None, 128)               0
dropout_20 (Dropout)         (None, 128)               0
dense_51 (Dense)             (None, 256)               33024
batch_normalization_42 (Bat  (None, 256)               1024
chNormalization)
activation_42 (Activation)   (None, 256)               0
dropout_21 (Dropout)         (None, 256)               0
dense_52 (Dense)             (None, 512)               131584
batch_normalization_43 (Bat  (None, 512)               2048
chNormalization)
activation_43 (Activation)   (None, 512)               0
dropout_22 (Dropout)         (None, 512)               0
dense_53 (Dense)             (None, 256)               131328
batch_normalization_44 (Bat  (None, 256)               1024
chNormalization)
activation_44 (Activation)   (None, 256)               0
dropout_23 (Dropout)         (None, 256)               0
dense_54 (Dense)             (None, 128)               32896
batch_normalization_45 (Bat  (None, 128)               512
chNormalization)
activation_45 (Activation)   (None, 128)               0
dropout_24 (Dropout)         (None, 128)               0
dense_55 (Dense)             (None, 68)                8772
=================================================================
Total params: 347,076
Trainable params: 344,516
Non-trainable params: 2,560
```

**Figure 8: NN-based architecture of BA classifier.**

```
Layer (type)                 Output Shape              Param #
=================================================================
dense (Dense)                (None, 64)                3648
batch_normalization (BatchN  (None, 64)                256
ormalization)
activation (Activation)      (None, 64)                0
dropout (Dropout)            (None, 64)                0
dense_1 (Dense)              (None, 128)               8320
batch_normalization_1 (Batc  (None, 128)               512
hNormalization)
activation_1 (Activation)    (None, 128)               0
dropout_1 (Dropout)          (None, 128)               0
dense_2 (Dense)              (None, 64)                8256
batch_normalization_2 (Batc  (None, 64)                256
hNormalization)
activation_2 (Activation)    (None, 64)                0
dropout_2 (Dropout)          (None, 64)                0
dense_3 (Dense)              (None, 155)               10075
=================================================================
Total params: 31,323
Trainable params: 30,811
Non-trainable params: 512
```

**Figure 9: NN-based architecture of privacy-preserving BA classifier.**

```
Layer (type)                 Output Shape              Param #
=================================================================
dense_30 (Dense)             (None, 128)               3968
batch_normalization_24 (Bat  (None, 128)               512
chNormalization)
activation_24 (Activation)   (None, 128)               0
dense_31 (Dense)             (None, 256)               33024
batch_normalization_25 (Bat  (None, 256)               1024
chNormalization)
activation_25 (Activation)   (None, 256)               0
dense_32 (Dense)             (None, 256)               65792
batch_normalization_26 (Bat  (None, 256)               1024
chNormalization)
activation_26 (Activation)   (None, 256)               0
dense_33 (Dense)             (None, 128)               32896
batch_normalization_27 (Bat  (None, 128)               512
chNormalization)
activation_27 (Activation)   (None, 128)               0
dense_34 (Dense)             (None, 33)                4257
=================================================================
Total params: 143,009
Trainable params: 141,473
Non-trainable params: 1,536
```

**Figure 10: NN-based architecture of attack model.**

[24] Valeria Nikolaenko, Udi Weinsberg, Stratis Ioannidis, Marc Joye, Dan Boneh, and Nina Taft. 2013. Privacy-preserving ridge regression on hundreds of millions of records. In *2013 IEEE symposium on security and privacy*. IEEE, 334–348.
[25] Olga Ohrimenko, Felix Schuster, Cédric Fournet, Aastha Mehta, Sebastian Nowozin, Kapil Vaswani, and Manuel Costa. 2016. Oblivious {Multi-Party} machine learning on trusted processors. In *25th USENIX Security Symposium (USENIX Security 16)*. 619–636.
[26] Seyed Ali Osia, Ali Shahin Shamsabadi, Ali Taheri, Kleomenis Katevas, Hamid R Rabiee, Nicholas D Lane, and Hamed Haddadi. 2017. Privacy-preserving deep inference for rich user data on the cloud. *arXiv preprint arXiv:1710.01727* (2017).
[27] Bhavesh Pandya, Georgina Cosma, Ali A Alani, Aboozar Taherkhani, Vinayak Bharadi, and TM McGinnity. 2018. Fingerprint classification using a deep convolutional neural network. In *2018 4th International Conference on Information Management (ICIM)*. IEEE, 86–91.
[28] Florian Schroff, Dmitry Kalenichenko, and James Philbin. 2015. Facenet: A unified embedding for face recognition and clustering. In *Proceedings of the IEEE conference on computer vision and pattern recognition*. 815–823.
[29] Elaine Shi, Yuan Niu, Markus Jakobsson, and Richard Chow. 2011. Implicit authentication through learning user behavior. In *Proceedings of ISC'2010*. Springer, 99–113.
[30] Somayeh Taheri, Md Morshedul Islam, and Reihaneh Safavi-Naini. 2017. Privacy-Enhanced Profile-Based Authentication Using Sparse Random Projection. In *Proceedings of the IFIP SEC'17*. Springer, 474–490.

[31] Somayeh Taheri, Md Morshedul Islam, and Reihaneh Safavi-Naini. 2017. Privacy-Enhanced Profile-Based Authentication Using Sparse Random Projection. In *IFIP International Conference on ICT Systems Security and Privacy Protection*. Springer, 474–490.
[32] Yongjin Wang et al. 2010. An analysis of random projection for changeable and privacy-preserving biometric verification. *IEEE Trans. on Systems, Man, and Cybernetics, Part B (Cybernetics)* 40 (2010), 1280–1293.
[33] Nan Zheng, Aaron Paloski, and Haining Wang. 2011. An efficient user verification system via mouse movements. In *Proceedings of the 18th ACM conference on Computer and communications security (CCS '11)*. ACM, 139–150.

## A APPENDICES

Figure 8 shows the architecture of the NN-based BA classifier.

## B APPENDICES

Figure 9 shows the architecture of the NN-based BA privacy-preserving classifier.

## C APPENDICES

Figure 10 shows the architecture of the NN-based attack model.